\documentclass[11pt]{article}

\usepackage{cite}

\usepackage{amsfonts}
\usepackage{amsthm}

\usepackage{amsmath}
\usepackage{amssymb}
\usepackage{enumerate}
\usepackage{url}
 
\newtheorem{theorem}{Theorem}
\newtheorem{lemma}{Lemma}
\newtheorem{remark}{Remark}

\newtheorem{proposition}{Proposition}

\newcommand{\E}{\mathbb{E}}
\newcommand{\Var}{\mathrm{Var}}

\newcommand{\R}{\mathbb{R}}

\newcommand{\Hess}{\operatorname{Hess}}

\newcommand{\EE}{\mathbb{E}}
\newcommand{\cI}{\mathcal{I}}

\newcommand{\EFI}{\E_\pi \mathcal{I}(\theta)}

\newcommand{\VP}{\Var(\pi)}

\usepackage{fullpage}
\title{A Family of Bayesian Cram\'er-Rao Bounds, \\
and Consequences for  Log-Concave Priors}
\author{Efe Aras, Kuan-Yun Lee, Ashwin Pananjady and Thomas~A.~Courtade\\University of California, Berkeley}
\date{February 22, 2019}

\begin{document}
\maketitle

\begin{abstract}
Under minimal regularity assumptions, we establish a family of information-theoretic Bayesian Cram\'er-Rao bounds, indexed by probability measures that satisfy a logarithmic Sobolev inequality.
This family includes as a special case the known Bayesian Cram\'er-Rao bound (or van Trees inequality), and its less widely known entropic improvement due to Efroimovich.  For the  setting of a log-concave prior,  we obtain a Bayesian Cram\'er-Rao bound which holds for any (possibly biased) estimator and, unlike the van Trees inequality,  does not depend on the Fisher information of the prior.   
\end{abstract}

\section{Introduction}
Throughout, we let $\mathcal{P}(\R^n)$ denote the set of  Borel probability measures on $\R^n$.  For $\mu\in\mathcal{P}(\R^n)$, we abuse notation slightly and define $\Var(\mu):=\inf_{c\in \R^n} \int|x-c|^2d\mu$, where $|\cdot|$ denotes Euclidean length on $\R^n$.  Thus, $\Var(\mu)$ is the usual variance in dimension $n=1$; it is the trace of the covariance matrix corresponding to $\mu$ for arbitrary dimension $n$.   A probability measure $\mu \in\mathcal{P}(\R^n)$ is said to be log-concave if $d\mu(x) = e^{-V(x)}dx$ for convex $V$.      All logarithms are taken with respect to the natural base.

Our results are best stated within the general framework of parametric statistics.  To this end, we let $(\mathcal{X},\mathcal{F},P_{\theta}; \theta \in \R^n )$ be a dominated family of probability measures on a measurable space $(\mathcal{X},\mathcal{F})$; with  dominating $\sigma$-finite measure  $\lambda$.   To each $P_{\theta}$, we associate a density $f(\,\cdot\,; \theta)$ (w.r.t.~$\lambda$) according to
\begin{align*}
dP_{\theta}(x) = f(x;\theta) d\lambda(x).
\end{align*}
The Fisher information of the parametric family $(P_{\theta})$ evaluated at $\theta$ is defined  as
$$
\mathcal{I}(\theta) := \int_{\mathcal{X}} \frac{|\nabla_\theta f(x;\theta) |^2}{f(x;\theta)} d\lambda(x),
$$
where $\nabla_\theta$ denotes gradient with respect to $\theta$.    Note that $\mathcal{I}$ is distinct from the information theorist's Fisher information $\mathcal{J}$, defined as
$$
\mathcal{J}(\mu) := \int_{\R^n} \frac{|\nabla \varrho(\theta)|^2}{\varrho(\theta)}d\theta.
$$
for  a probability measure $\mu \in \mathcal{P}(\R^n)$ having density $\varrho$ with respect to Lebesgue measure.  In the special case where $\theta$ is a location parameter,  the two   quantities   coincide.   

For a real-valued parameter $\theta\in \R$ and an observation $X\sim P_{\theta}$, the basic question of parametric statistics is how well  can one estimate $\theta$ from $X$.  Here, the Cram\'er-Rao bound is of central importance in proving lower bounds on $L^2$ estimation error, stating that 
\begin{align}
\Var(\hat{\theta}(X)) = \EE(\theta-\hat{\theta}(X))^2\geq \frac{1}{\mathcal{I}(\theta)}\label{eq:CRB}
\end{align}
for any unbiased estimator $\hat{\theta}$.  The assumption of unbiasedness is quite restrictive, especially since unbiased estimators may not  always exist, or may be less attractive than biased estimators for any one of a variety of reasons (computability, performance, etc.).  Under the assumption that the parameter $\theta$ is distributed according to some prior $\pi\in \mathcal{P}(\R)$, the so-called Bayesian Cram\'er-Rao bound \cite{gill1995applications, van2004detection} (also known as the van Trees inequality) states, under mild regularity assumptions,  that
\begin{align}
	\E(\theta-\hat{\theta}(X))^2 \geq \frac{1}{\EFI + \mathcal{J}(\pi)},
	\label{eq:van-Trees inequality}
\end{align}
where the expectation is over $\theta \sim \pi$ and, conditioned on $\theta$,  $X\sim P_{\theta}$.  As noted by Tsybakov \cite[Section 2.7.3]{Tsybakov}, this inequality is quite powerful since it does not impose any restriction on unbiasedness, is relatively simple to apply, and often leads to sharp results (including sharp constants).   Tsybakov    states that one primary disadvantage of \eqref{eq:van-Trees inequality} is that it applies only to $L^2$ loss.   Although it does not appear to be widely known, this is actually not true.  Indeed, Efroimovich proved in \cite{efroimovich1979information} that
\begin{align}
	\frac{1}{2\pi e}e^{2 h(\theta|X)} \geq \frac{1}{\EFI + \mathcal{J}(\pi)},
	\label{eq:EfroimovichInequality}
\end{align}
which is  stronger than \eqref{eq:van-Trees inequality} by the maximum-entropy property of Gaussians.   Efroimovich's inequality can be rearranged to give an upper bound on the mutual information 
\begin{align*}
I(\pi; P_{\theta}) &\equiv I(\theta; X) := \iint f(x;\theta) \log \frac{f(x;\theta)}{\int  f(x;\theta') d\pi(\theta')} d\lambda(x) d\pi(\theta).
\end{align*} Such a general upper bound on $I(\pi;P_{\theta})$ can be useful in settings beyond those where \eqref{eq:van-Trees inequality} applies.  For example, it can be used to give one direction of the key estimate in Clarke and Barron's  work showing that Jeffrey's prior is least favorable \cite{clarke1994jeffreys}.   It  can also be applied to characterize Bayes risk measured under losses other than $L^2$ when coupled with a lower bound on mutual information (see, e.g., \cite{YihonLN}). We remark that several systematic techniques exist for lower bounding the mutual information $I(\pi; P_{\theta})$ in terms of Bayes risk (e.g., Fano's method, or the Shannon lower bound for the rate distortion function),  so finding a good upper bound is often the challenge.  A typical heuristic is to bound $I(\pi; P_{\theta})$ from above by the capacity of the channel $\theta \mapsto P_{\theta}$, but this method has the disadvantages that  (i) it discards information about the prior $\pi$; and 
(ii) capacity expressions are only explicitly known for very special parametric families $(P_{\theta})$ (e.g., Gaussian channels).   Efroimovich's inequality overcomes both of these obstacles, but has the undesirable property of being degenerate when $ \mathcal{J}(\pi)=+\infty$.  This can be a serious disadvantage in applications since many natural  priors   have infinite Fisher information, for example uniform measures on convex bodies\footnote{Mollification may be a useful heuristic to compensate for infinite $\mathcal{J}(\pi)$ in low dimensions, but this  becomes  fundamentally problematic in high dimensions where mollification picks up dimensional dependence, and generally alters the boundary of a set where the measure concentrates.}.  

\subsection*{Contributions}

We make two  main contributions, which we describe in rough terms here.  Precise statements are given in Section \ref{sec:MainResults}.    First, we establish a family of Bayesian Cram\'er-Rao-type bounds indexed by probability measures that satisfy a logarithmic Sobolev inequality on $\R^n$.  This generalizes Efroimovich's inequality \eqref{eq:EfroimovichInequality}, which  corresponds to the special case where the reference measure is taken to be Gaussian.  Second, we specialize the first result to obtain an explicit Bayesian Cram\'er-Rao-type bound under the assumption of a log-concave prior $\pi$. In dimension one, 
the result implies 
\begin{align}
e^{2 h(\theta|X)}\geq \frac{4}{e^2 \EFI } \approx  \frac{0.54}{ \EFI },    \label{eqBCRBoned}
\end{align}
provided $\Var(\pi)\geq 1/\EFI$; a  correction is needed if this condition is not met\footnote{It is easy to see why a condition like this is needed:  if there were no such assumption, then we could let $\pi$  approximate a point mass, effectively showing that the Cram\'er-Rao bound holds -- up to an absolute constant -- for \emph{any} estimator.  This  clearly can not be true (consider $\hat\theta$ constant, not equal to $\theta$). } (see Theorem \ref{thm:CRB-logconcave} for a precise statement).  In particular, 
$$\EE(\theta-\hat{\theta}(X))^2\geq \Var(\theta-\hat{\theta}(X)) \geq  C\frac{1}{ \EFI }$$
holds under our assumptions for a universal constant $C\geq 4 e^{-2}\approx 0.54$, regardless of whether $\hat{\theta}$ is biased.  This should be compared to the classical Cram\'er-Rao bound: morally speaking,   \eqref{eq:CRB} continues to hold (up to a modest constant factor) for any estimator $\hat{\theta}$, provided we are working with a log-concave prior $\pi$ which, together with $(P_{\theta})$, satisfies $\Var(\pi)\geq 1/\EFI$.   
Note that the crucial (and somewhat surprising) advantage relative to  \eqref{eq:EfroimovichInequality} is that the Fisher information $\mathcal{J}(\pi)$ does not appear.  

\subsection*{Organization}
The sequel is organized as follows: main results, along with assumptions and brief discussion are provided in Section \ref{sec:MainResults}.  The proofs of all results can be found in Section \ref{sec:outline-of-proofs}.

\section{Main Results}\label{sec:MainResults}
\subsection{Assumptions}
As is typical of Cram\'er-Rao-type bounds, our main results require us to assume some mild regularity.  In particular, for a given measure $\mu\in \mathcal{P}(\R^n)$, we will  refer to the following standard condition on the densities associated to $(P_{\theta})$: 
\begin{align}
\int_{\mathcal{X}} \nabla_\theta f(x;\theta) d\lambda(x) = 0, \hspace{1cm}\mu-a.e.~\theta, \label{assum:Reg}
\end{align}
where $\nabla_\theta$ denotes the gradient with respect to $\theta$.  We remark that this   holds whenever the orders of differentiation with respect to $\theta$ and integration with respect to $x$ can be exchanged (Liebniz rule).

\subsection{Statement of Results}
Our first main result establishes a family of Cram\'er-Rao-type bounds on the mutual information $I(\pi;P_{\theta})$ in terms of logarithmic Sobolev inequalities on $\mathbb{R}^n$.  To this end, we  recall the standard definitions of relative entropy and relative Fisher information (the parlance in which logarithmic Sobolev inequalities are framed).  Consider $\mu, \nu \in \mathcal{P}(\R^n)$, with $\nu \ll \mu$ and $d\nu = h d\mu$. The  entropy of $\nu$, relative to $\mu$, is defined as
$$
D_{\mu}(\nu)  \equiv D_{\mu}(h)   := \int_{\R^n} h \log h d\mu.
$$
If the density $h$ is weakly differentiable, the Fisher information of $\nu$, relative to $\mu$, is defined according to 
$$
I_{\mu}(\nu)  \equiv I_{\mu}(h)     := \int_{\R^n} \frac{|\nabla h|^2}{h}d\mu.
$$
If $h$ is not weakly differentiable, we adopt the convention that $I_{\mu}(h)=+\infty$ so that our expressions make sense even in the general case.

A probability measure $\mu$ is said to satisfy a logarithmic Sobolev inequality with constant $C>0$ (or, $\mathrm{LSI}(C)$ for short) if, for all probability measures $\nu\ll \mu$, 
$$
D_{\mu}(\nu)\leq \frac{C}{2 } I_{\mu}(\nu).
$$
The standard Gaussian measure $d\gamma(x) := (2\pi)^{-n/2} e^{-|x|^2/2}dx$ on $\R^n$ is a prototypical example of a measure that satisfies an LSI, and does so with constant $C=1$.  More generally, if $d\mu(x) = e^{-V(x)} dx$ with $\Hess(V)\geq K \cdot \mathrm{I}_{n}$ for $K>0$ and $\mathrm{I}_{n}$ the $n\times n$ identity matrix, then $\mu$ satisfies  $\mathrm{LSI}(1/K)$ \cite{bakry1985diffusions}; this result is known as the Bakry-\'Emery theorem, and we shall need it later in the proof of Theorem \ref{thm:CRB-logconcave}.  

With these definitions in hand, our first result is the following:
\begin{theorem}\label{thm:BayesianLSI}
Let $\mu \in \mathcal{P}(\mathbb{R}^n)$ satisfy $\mathrm{LSI}(C)$ and assume the regularity condition~\eqref{assum:Reg} holds. For any probability measure $\pi \ll\mu$ on $\mathbb{R}^n$,
\begin{equation}
    I(\pi;P_\theta) + D_\mu(\pi) \leq \frac{C}{2} \left(I_\mu(\pi) + \int_{\mathbb{R}^n} \mathcal{I}(\theta) d \pi(\theta) \right).
    \label{eq:thm-1-eq}
\end{equation}
\label{thm:thm-1}
\end{theorem}
Inequality \eqref{eq:thm-1-eq}  improves the LSI for $\mu$.  Indeed, taking $P_{\theta}$ independent of $\theta$ renders $I(\pi;P_\theta) =\mathcal{I}(\theta)=0$, so that the LSI for $\mu$ is recovered. However, the proof of  \eqref{eq:thm-1-eq}  follows from a relatively simple application of the LSI for $\mu$ and some basic calculus, so the two inequalities should be viewed as being formally equivalent in this sense.  

Clearly, the statement of Theorem \ref{thm:BayesianLSI} allows us the freedom to choose the measure $\mu$ so as to obtain the tightest possible bound on $I(\pi;P_\theta)$.  However, a notable example is obtained when $\mu$ is taken to be the standard Gaussian measure on $\R^n$.  In this case, upon simplification we obtain 
\begin{equation}
1  +  \log(2\pi e)   \leq \frac{2}{n}h(\theta|X)  +    \mathcal{J}(\pi) + \int_{\R^n} \mathcal{I}(\theta) d\pi(\theta) .
	\label{eq:efroimovitch}
\end{equation}
Of note, \eqref{eq:efroimovitch} is not invariant to rescalings of the parameter $\theta$.  So, just as one passes from  Lieb's inequality to the entropy power inequality, we may optimize over all such scalings to obtain the following multidimensional version of \eqref{eq:EfroimovichInequality}:
$$
\frac{1}{2\pi e}\exp\left( \frac{2}{n}h(\theta|X) \right)\geq \frac{n}{  \mathcal{J}(\pi) + \int_{\R^n} \mathcal{I}(\theta) d\pi(\theta) }.
$$
\begin{remark}
Efroimovich's work \cite{efroimovich1979information} contains a slightly stronger multidimensional form, stated in terms of determinants of Fisher information matrices.  As defined, our Fisher information quantities $\mathcal{I}$ and $\mathcal{J}$ correspond to traces of the same matrices, leading to a weaker inequality by the arithmetic-geometric mean inequality.   Nevertheless, the two inequalities should really be regarded as essentially equivalent, as they are both  direct consequences of the one-dimensional inequality (where the two results coincide).  See \cite[Proof of Theorem 5]{efroimovich1979information} for details.  It is unclear whether a similar claim holds for non-Gaussian $\mu$ in \eqref{eq:thm-1-eq}.
\end{remark}

We remark that \eqref{eq:EfroimovichInequality} was discovered by Efroimovich in 1979, but does not appear to be widely known (we could not find a statement of the result outside the Russian literature).  At the time of  Efroimovich's initial discovery of  \eqref{eq:EfroimovichInequality}, the study of logarithmic Sobolev inequalities was just getting started, being largely initiated by Gross's work on the Gaussian case in 1975 \cite{gross1975logarithmic}.  In particular, the derivation of  \eqref{eq:EfroimovichInequality} (and, less generally, the van Trees inequality) from the Gaussian  logarithmic Sobolev inequality does not appear to have been observed previously.  So, from a conceptual standpoint, one contribution of Theorem \ref{thm:BayesianLSI} is that it demonstrates how Efroimovich's result (and the weaker  van Trees inequality) emerges as one particular instance in the broader context of LSIs which, to our knowledge, have not found direct use in parametric statistics beyond their implications for measure concentration (see, e.g., \cite{ledoux2001concentration}).

A nontrivial consequence of Theorem \ref{thm:BayesianLSI} is a general Cram\'er-Rao-type bound on $I(\pi; P_{\theta})$, assuming only that 
$\pi$ is log-concave.   Specifically, our second main result is the following:
 \begin{theorem}\label{thm:CRB-logconcave}
Assume  the parametric family $(P_{\theta})$ satisfies \eqref{assum:Reg} for $\mu$ equal to Lebesgue measure.     Let $d\pi(x) = e^{-V(x)}dx$ satisfy    $\Hess(V) \geq K\cdot \mathrm{I}_n$ for some scalar $K\geq 0$, where  $\mathrm{I}_n$ is the $n\times n$ identity matrix.   Define $P := \frac{1}{n} \VP$, $J:= \frac{1}{n} \int_{\mathbb{R}^n}\mathcal{I}(\theta)d\pi(\theta)$.  It holds that
    \begin{align}
        I(\pi;P_\theta) \leq n\cdot \phi\left( \sqrt{(K P)^2+ J  P} - KP\right),\label{eq:logConcaveCRB}
    \end{align}
    where
    \begin{align*}
    \phi(x):=
    \begin{cases}
    x &\mbox{if $0\leq x < 1$ }\\
    1+\log x  & \mbox{if $x\geq 1$. }
    \end{cases}
    \end{align*}
\end{theorem}
\begin{remark}
The one-dimensional inequality \eqref{eqBCRBoned} follows directly from Theorem \ref{thm:CRB-logconcave} for $K=0$, combined with the entropy lower bound for log-concave random variables $h(\theta)\geq \frac{1}{2}\log(4 \Var(\theta))$ due to Marsiglietti and Kostina \cite{marsiglietti2018lower}.   Similar statements hold for general dimension $n$, albeit with a correction factor that depends on dimension (no correction is needed if the hyperplane conjecture is true; see \cite{bobkov2011entropy}). 
\end{remark}
 
The upper bound \eqref{eq:logConcaveCRB} should be viewed as a function of two nonnegative quantities: the products $KP$ and $JP$.  By the Brascamp-Lieb inequality \cite{brascamp1976extensions}, we always have $KP\leq 1$; this quantity only depends on the prior $\pi$ and distills what quantitative information is known about its degree of log-concavity.  In particular, if $\pi$ is only known to be log-concave, then $K=0$ gives  $I(\pi;P_\theta) \leq n\cdot \phi\left( \sqrt{  J  P}  \right)$.  In the other extreme case, if $KP=1$ (e.g., if $\pi$ is scaled standard Gaussian), we have the slightly improved bound $ I(\pi;P_\theta) \leq n\cdot \phi\left( \sqrt{ 1+ J  P}  -1\right)$.  These bounds both essentially behave as $\frac{n}{2}\log(JP)$ for $JP$ modestly large, so knowledge of $KP$ (i.e., additional  information about the measure $\pi$) only significantly affects the behavior of the upper bound  \eqref{eq:logConcaveCRB}  for $JP$ small.  To be precise, for $JP$  near zero, the upper bound behaves as $n J/K$ when $K>0$, and $n \sqrt{JP}$ if $K=0$.   Applications in asymptotic statistics   consider a sequence of observations $X_1, \dots, X_m$, conditionally independent given $\theta$.  In this case, $J$ grows linearly with $m$, so that the logarithmic behavior of the bound dominates, regardless of what is known about $K$.

Let us now make a brief observation on the sharpness of  Theorem \ref{thm:CRB-logconcave}. To this end, consider the classical Gaussian sequence model $X = \theta + Z$, where $Z\sim N(0,\sigma^2 \mathrm{I}_{n})$ is independent of $\theta \sim \pi$.  In this case, the 
typical  quantity of relevance is the signal-to-noise ratio $\mathsf{snr} := \frac{\VP}{n \sigma^2} =n^{-2}  \VP    \int \mathcal{I}(\theta)d\pi(\theta)$, in terms of which   we have the sharp upper bound
\begin{align}
I(\pi; P_\theta) \leq \frac{n}{2}\log(1+\mathsf{snr}) = \frac{n}{2}\log(1+JP) . \label{eq:gsm}
\end{align}
Thus, in view of the previous discussion, we clearly see that Theorem \ref{thm:CRB-logconcave} provides a  sharp estimate in the regime where $JP$ is moderately large.   We do not yet know whether the bound $I(\pi;P_\theta) \leq n\cdot \phi\left( \sqrt{  J  P}  \right)$ is sharp for small $JP$ and $K=0$, but we believe that it should be.   

Finally, we remark that all results have correct dependence on dimension, as can be seen by testing on product measures. 

\subsection{Remarks on Applications}
Applications of Cram\'er-Rao-type bounds to parameter estimation are  numerous, and our results will generally apply in Bayesian settings.  In particular, we believe corollaries such as \eqref{eqBCRBoned} may be especially useful for proving lower bounds on Bayes risk when the prior $\pi$ is  log-concave.   

We note that our results are quite general in form, and therefore not restricted to applications in parametric statistics.  To give one quick example, consider log-concave $\mu\in \mathcal{P}(\R^n)$, normalized so that $\Var(\mu) = n$,  and define $S_k = \sum_{i=1}^k X_i$, where $X_i$ are drawn i.i.d. according to $\mu$.  Then, an immediate corollary of Theorem \ref{thm:CRB-logconcave} is that, for $k$ sufficiently large, 
$$
\exp\left( {\frac{2}{n} h(S_k)} \right)\leq \left( k e^2 \frac{\mathcal{J}(\mu)}{n}\right) \exp\left( {\frac{2}{n} h(S_1)}\right),
$$
which is a sort of reverse entropy power inequality, holding for log-concave random vectors.    This improves a result of Cover and Zhang \cite{cover1994maximum} for $k$ sufficiently large, in which the leading coefficient in parentheses on the right is $k^2$.  This inequality should also be compared to the formulation of the hyperplane conjecture recently put forth by Marsiglietti and Kostina \cite{marsiglietti2018new}.

\section{Proofs}
\label{sec:outline-of-proofs}
This section contains the proofs of main results.  
\subsection{Proof of Theorem \ref{thm:thm-1}}
\label{subsec:proof-of-thm1}
We may assume that the RHS of equation~\eqref{eq:thm-1-eq} is finite; else the claim is trivially true.  Let $d\pi = h d\mu$, and note that $h(\theta) f(x;\theta)$ is the joint density of $(\pi, P_{\theta})$ with respect to $\mu\times \lambda$.   Define $f(x) = \int_{\R^n} f(x;\theta) d\pi(\theta)$, and $h_x(\theta) = h(\theta) f(x;\theta)/f(x)$, which is well-defined $(\pi\times \lambda)$-a.e. Now, since $\mu$ satisfies $\mathrm{LSI}(C)$, we have for $\lambda$-a.e.~$x$
\begin{align*}
\int_{\R^n} h_x(\theta) \log h_x(\theta) d\mu(\theta) \leq \frac{C}{2} \int_{\R^n} \frac{|\nabla h_x(\theta)|^2}{h_x(\theta)}d\mu(\theta),
\end{align*}
where we write $\nabla$ in place of $\nabla_{\theta}$ for brevity. 
Integrating both sides with respect to the density $f d\lambda$, we have 
\begin{align*}
&\int_{\mathcal{X}} f(x)\left( \int_{\R^n} h_x(\theta) \log h_x(\theta) d\mu(\theta) \right)d\lambda(x)\leq \frac{C}{2} \int_{\mathcal{X}} f(x)\left( \int_{\R^n} \frac{|\nabla h_x(\theta)|^2}{h_x(\theta)}d\mu(\theta)\right)d\lambda(x).
\end{align*}
Now, observe that 
\begin{align*}
&\int_{\mathcal{X}} f(x)\left( \int_{\R^n} \frac{|\nabla h_x(\theta)|^2}{h_x(\theta)}d\mu(\theta)\right)d\lambda(x)\\
&=\int_{\mathcal{X}}  \int_{\R^n} \frac{|\nabla (f(x) h_x(\theta))|^2}{f(x) h_x(\theta)}d\mu(\theta) d\lambda(x)\\
&=\int_{\mathcal{X}}  \int_{\R^n} \frac{|\nabla (f(x;\theta) h(\theta))|^2}{f(x;\theta) h(\theta)}d\mu(\theta) d\lambda(x)\\
&=\int_{\mathcal{X}}  \int_{\R^n} \left( f(x;\theta) \frac{|\nabla h(\theta)|^2}{h(\theta)}   + 2 \nabla h(\theta) \cdot \nabla f(x;\theta) +  h(\theta) \frac{|\nabla f(x;\theta)|^2}{f(x;\theta)}     \right) d\mu(\theta) d\lambda(x)\\
&= I_{\mu}(\pi)  +   \int_{\R^n} \cI(\theta) d\pi(\theta)  + 2 \int_{\mathcal{X}}  \int_{\R^n} \nabla h(\theta) \cdot \nabla f(x;\theta) ,
\end{align*}
where the penultimate identity follows by the product rule for derivatives and expanding the square.  
The final cross term is integrable; indeed,  Cauchy-Schwarz   yields
\begin{align*}
&\int_{\mathcal{X}} \int_{\R^n} |\nabla h(\theta) \cdot \nabla f(x;\theta) | d\mu(\theta) d\lambda(x)\\
&\leq \sum_{i=1}^d \int_{\mathcal{X}}  \int_{\R^n} |\partial_{\theta_i} h(\theta)  \partial_{\theta_i}  f(x;\theta) d\mu(\theta) |d\lambda(x) \\
&\leq \sum_{i=1}^d\left( \int_{\mathcal{X}}  \int_{\R^n}\frac{|\partial_{\theta_i} h(\theta)|^2}{h(\theta)}f(x;\theta)    d\mu(\theta) d\lambda(x)  \right)^{1/2} \left( \int_{\mathcal{X}}  \int_{\R^n}\frac{|\partial_{\theta_i} f(x;\theta)|^2}{f(x;\theta)}h(\theta)  d\mu(\theta) d\lambda(x)  \right)^{1/2}\\
&\leq \sqrt{  I_{\mu}(\pi)     \int_{\R^n} \cI(\theta) d\pi(\theta) }.
\end{align*}
The exchange of integrals to obtain the last line is justified by Tonelli's theorem.   Therefore, by Fubini's theorem, 
\begin{align*}
\int_{\mathcal{X}}  \int_{\R^n} \nabla h(\theta) \cdot \nabla f(x;\theta)  d\mu(\theta) d\lambda(x)   =  \int_{\R^n} \nabla h(\theta) \cdot    \left( \int_{\mathcal{X}}    \nabla f(x;\theta)  d\lambda(x) \right) d\mu(\theta)  =0,
\end{align*}
where the last equality follows by the regularity assumption.  Summarizing, we have
\begin{align*}
&\int_{\mathcal{X}} f(x)\left( \int_{\R^n} \frac{|\nabla h_x(\theta)|^2}{h_x(\theta)}d\mu(\theta)\right)d\lambda(x) =  I_{\mu}(\pi)  +   \int_{\R^n} \cI(\theta) d\pi(\theta).
\end{align*}
To finish, we observe that 
\begin{align*}
&\int_{\mathcal{X}} f(x)\left( \int_{\R^n} h_x(\theta) \log h_x(\theta) d\mu(\theta) \right)d\lambda(x) \\
&=\int_{\mathcal{X}}   \int_{\R^n} f(x) h_x(\theta) \log h_x(\theta) d\mu(\theta)  d\lambda(x)\\
&=\int_{\mathcal{X}}   \int_{\R^n}  f(x;\theta) h(\theta) \log \frac{h_x(\theta)}{h(\theta)} d\mu(\theta)  d\lambda(x) + \int_{\mathcal{X}}   \int_{\R^n} f(x;\theta) h(\theta) \log h(\theta) d\mu(\theta)  d\lambda(x)\\
&=\int_{\mathcal{X}}   \int_{\R^n}  f(x;\theta) h(\theta) \log \frac{f(x;\theta)}{f(x)} d\mu(\theta)  d\lambda(x) + \int_{\mathcal{X}}   \int_{\R^n}  h(\theta) \log h(\theta) d\mu(\theta)\\
&= I(\pi; P_{\theta}) + D_{\mu}(\pi),
\end{align*}
which proves the claim. 
\subsection{Proof of Theorem \ref{thm:CRB-logconcave}}
\label{subsec:proof-of-thm23}
We require the following  proposition, the proof of which is the most arduous part of the argument.  The ideas of the proof are  independent from Theorem \ref{thm:CRB-logconcave}, so it is deferred to  the appendix. 
\begin{proposition} \label{prop:CdeltaLemma}Let $\rho = e^{-V}$ be a probability density on $\R^n$, with $V$ convex. 
\begin{enumerate}
\item[(i)]
 For each $\delta> 0$, there exists a unique $m_{\delta} \in \R^n$ such that 
\begin{align*}
\hspace{-0.7cm}\int_{\R^n} x e^{-\delta |x-m_{\delta} |^2/2}\rho(x) dx = m_{\delta}   \int_{\R^n} e^{-\delta |x-m_{\delta} |^2/2}\rho(x) dx  . 
\end{align*}
\item[(ii)]   For $m_{\delta}$ as in part (i), and each $\delta\geq 0$
\begin{align*}
 -\log\left( \int_{\R^n} e^{- \delta |x-m_{\delta}|^2/2  } \rho(x)dx\right) \leq 
\begin{cases}
\frac{\delta}{2}\Var(\rho)  & \mathrm{if}~  0 \leq   \delta  < \frac{n}{ \Var(\rho)},\\
\frac{n}{2}\left(1 + \log \left(\frac{\delta}{n}  \Var(\rho)  \right)\right) &    \mathrm{if} ~ \delta  \geq \frac{n}{\Var(\rho)}.
\end{cases}
\end{align*}

\end{enumerate}
\label{lem:lem-bcrb-4}
\end{proposition}

 To begin the proof, consider the  log-concave density $d\pi(x) = e^{-V(x)}dx$, where  $\operatorname{Hess}(V)\geq K\cdot \mathrm{I}_n$. For $\delta>0$, let $\mu_\delta$ be the probability measure with density
$$
d\mu_\delta(x) = C_\delta^{-1} e^{-V(x)-\delta|x-m_\delta|^2/2}dx,
$$
where $C_\delta = \int e^{-V(x)-\delta|x-m_\delta|^2/2}dx$ is a normalizing constant and $m_{\delta} \in \R^n$ is such that $
\int_{\R^n} x d\mu_{\delta} = m_{\delta}$, which exists as a consequence of Proposition \ref{prop:CdeltaLemma}(i).    Note that  $\pi$ has density $C_{\delta} e^{ \delta | x- m_{\delta}|^2/2  }$ with respect to $\mu_\delta$.  Therefore, we may readily compute
\begin{align*}
D_{\mu_\delta }(\pi)  &= \frac{\delta}{2}\int_{\R^n}    | x-m_{\delta}|^2  e^{-V(x)}dx  + \log C_{\delta}  = \frac{1}{2\delta} I_{\mu_\delta }(\pi )  + \log C_{\delta}.
\end{align*}
By the Bakry-Emery theorem, $\mu_\delta$ satisfies $\mathrm{LSI}(1/(K+\delta))$, so it follows from Theorem \ref{thm:thm-1} that 
\begin{align*}
I(\pi;P_\theta) &\leq -D_{\mu_{\delta}}(\pi) + \frac{1}{2(K+\delta)}\cdot I_{\mu_{\delta}}(\pi) + \frac{1}{2(K+\delta)} \int \mathcal{I}(\theta) d\pi(\theta) \\
&= -\frac{K}{2\delta(K+ \delta)} \cdot I_{\mu_{\delta}}(\pi) + \frac{1}{2(K+\delta)} \int \mathcal{I}(\theta) d\pi(\theta) - \log C_{\delta} \\
&=  -\frac{K\delta}{2 (K+{\delta})} \cdot \int |x-m_\delta |^2 e^{-V(x)} dx + \frac{1}{2(K+\delta)} \int \mathcal{I}(\theta) d\pi(\theta) - \log C_{\delta} .
\end{align*} 
By Proposition \ref{prop:CdeltaLemma}(ii) and the inequality 
$$\int |x-m_\delta |^2 e^{-V(x)} dx \geq \Var(\pi)$$ holding by definition of $\Var(\pi)$, we have
\begin{align}
    I(\pi; P_\theta) &\leq -\frac{K\delta }{2 (K+\delta)} \cdot nP + \frac{1}{2(K+\delta)} \cdot nJ  +\begin{cases}
            \frac{\delta}{2} \cdot nP & \text{if $0\leq \delta < \frac{1}{P}$} \\
            \frac{n}{2}\left(1 + \log \left( \delta P\right) \right) & \text{if $\delta \geq \frac{1}{P},$}
       \end{cases}
    \label{eq:main-proof-eq}
\end{align}
where $J,P$ are as defined in the statement of the theorem.  
Since the above holds for arbitrary $\delta>0$, we now particularize by (optimally) choosing 
$$
\delta = \sqrt{K^2+J/P}-K    
$$
if $JP < 1+2KP$, and otherwise choosing 
$$
\delta = \tfrac{1}{2} \left((K^2P + J - 2K) + \sqrt{(K^2P+J)^2 - 4K(K^2P+J)}\right) .
$$
It can be verified that if $JP < 1+2KP$, then this choice of $\delta$ ensures $\delta <1/P$.  On the other hand, if $JP \geq 1+2KP$, then this choice of $\delta$ ensures $\delta \geq1/P$.  Hence, substitution  into equation~\eqref{eq:main-proof-eq}  and simplifying yields:
    \begin{align*}
        I(\pi;P_\theta) \leq n\cdot \psi( K P, J  P)
    \end{align*}
    where $\psi$ is defined piecewise according to
   \begin{align*}
  \psi(a,b) := \begin{cases}
   \sqrt{a^2+b} - a  & \text{if $b < 2a + 1$}\\
    \frac{1}{2} \left(
                        1-a+\frac{2(a^2 + b)}{a^2 + b+\sqrt{(a^2+b)^2-4a(a^2+b)}}  + \log\left(\frac{a^2 + b+\sqrt{(a^2+b)^2-4a(a^2+b)}}{2} - a\right)
                    \right) & \text{otherwise.}
  \end{cases} 
   \end{align*}

   This bound is actually better than what is stated in the theorem, but is clearly a bit cumbersome.   Since $KP\leq 1$, we note the simpler (yet, still essentially as good) bound holding for $\psi$ in the range $0\leq a \leq 1$, completing the proof
\begin{align*}
\psi(a,b) \leq \begin{cases}
 \sqrt{a^2+b} - a  & \text{if $b < 2a + 1$}\\
 1 + \log\left( \sqrt{a^2+b} - a\right)  & \text{otherwise.}
\end{cases}
\end{align*}

\section*{Acknowledgement}
This work was supported  in part by NSF grants CCF-1704967, CCF-0939370 and CCF-1750430.

\bibliographystyle{unsrt}
\bibliography{BCRBbib}

\section*{Appendix}
 
This appendix contains the proof of the   following extended version of Proposition \ref{prop:CdeltaLemma}. It may be of independent interest. 
\begin{lemma} \label{lem:PropertiesOfLC}
Let $\rho = e^{-V}$ be a probability density on $\R^n$, with $V$ convex. 
\begin{enumerate}[(i)]
\item
 For each $\delta> 0$, there exists a unique $m_{\delta} \in \R^n$ such that 
\begin{align*}
\hspace{-0.7cm}\int_{\R^n} x e^{-\delta |x-m_{\delta} |^2/2}\rho(x) dx = m_{\delta}   \int_{\R^n} e^{-\delta |x-m_{\delta} |^2/2}\rho(x) dx  . 
\end{align*}

\item For each $\delta>0$,  the map
$$
m \longmapsto  \int_{\R^n} e^{-\delta |x- m |^2/2}\rho(x) dx 
$$
has a unique global maximum at $m_{\delta}$.

\item The map $\delta \longmapsto m_{\delta}$ is  continuous on $\delta \in (0,\infty)$.  In particular,  for each $\delta>0$, there is a neighborhood $U_{\delta}$ of $\delta$ and $L_{\delta}<\infty$ such that $|m_{\delta'}-m_{\delta}|\leq L_{\delta} |\delta'-\delta|$ for all $\delta'\in U_{\delta}$.  

\item For $m_{\delta}$ as in part (i), and each $\delta \geq 0$,
\begin{align*}
-\log\left( \int_{\R^n} e^{- \delta |x-m_{\delta}|^2/2  } \rho(x)dx\right)  \leq 
\begin{cases}
\frac{\delta}{2}\Var(\rho)  & \mathrm{if}~  0 \leq   \delta  < \frac{n}{ \Var(\rho)},\\
\frac{n}{2}\left(1 + \log \left(\frac{\delta}{n}  \Var(\rho)  \right)\right) &    \mathrm{if} ~ \delta  \geq \frac{n}{\Var(\rho)}.
\end{cases}
\end{align*}
\end{enumerate}
\end{lemma}
\begin{remark}
An intuitive interpretation is as follows: If we convolve a log-concave density with a Gaussian of variance $\delta^{-1}$, then the point of maximum likelihood of the resulting density (call it $m_{\delta}$) is unique, and changes smoothly as we adjust $\delta$.  The last part of the lemma gives a lower bound on the  likelihood at $m_{\delta}$. The only real surprise is the fact that $m_{\delta}$ is also the barycenter of the density proportional to $e^{- \delta |x-m_{\delta}|^2/2  } \rho(x)$, which is part (i) of the claim.
\end{remark}

The proof of Lemma \ref{lem:PropertiesOfLC} starts by showing that the map $T_{\delta}:\R^n\longrightarrow \R^n$ defined by
$$
T_{\delta} : m\longmapsto \frac{\int_{\R^n} x e^{-\delta |x-m|^2/2}\rho(x) dx  }{\int_{\R^n} e^{-\delta |x-m|^2/2}\rho(x) dx }
$$
is a contraction with respect to the usual Euclidean metric.   Then, the claims follow from the well-known Banach fixed-point theorem:
\begin{lemma}[Banach Fixed Point Theorem]\label{lem:BanachFixedPointTheorem}
Let $(X,d)$ be a complete metric space, and let $T:X\longrightarrow X$ satisfy $d(T(x),T(y))\leq \lambda d(x,y)$ for all $x,y\in X$, where $\lambda<1$.  Then $T$ has a unique fixed point $x^*\in X$.  Moreover, if $x_0\in X$ and  $x_{n+1} := T(x_n)$, $n\geq 0$, then 
\begin{equation}
d(x_n,x^*) \leq \frac{\lambda^n}{1-\lambda} d(T(x_0),x_0), \hspace{1cm} n\geq 0.
\end{equation}
\end{lemma}
So, to begin, let $\mu_{m,\delta}$ denote the probability measure with density proportional to $ e^{-\delta |x-m|^2/2}\rho(x)$. We note that $\mu_{m,\delta}$ cannot split off an independent Gaussian factor with variance $1/\delta$.  Indeed, if this were the case, then after suitable change of coordinates, we could assume $\mu_{m,\delta}$  splits off an independent Gaussian factor of variance $1/\delta$ in the first coordinate, so that
$$
e^{-V(x)-\delta |x-m|^2/2} \propto e^{-W(x_2, \dots, x_n) - \delta|x_1-c|^2/2 }
$$
for some $c\in \R$.  Rearranging, this yields $V(x) = W(x_2, \dots, x_n) + x_1(c-m_1) + C$ for some constant $C$.  
This would imply $\rho$ is not integrable in coordinate $x_1$, a contradiction.  Thus, we must have
$$
\sup_{\sigma \in S^{d-1}} \Var_{\mu_{m,\delta}}(x\mapsto \sigma \cdot x) \leq  \frac{\lambda_{\delta}}{\delta}
$$
for some $\lambda_{\delta}<1$. This follows from the Brascamp-Lieb inequality, and the fact that Gaussians are the only extremizers. 
 
By differentiating the $i$th coordinate of $T_{\delta}$ at $m$, we see that 
\begin{align*}
&\nabla [T_{\delta}]_{i}(m) \\
&=  \delta \frac{\int_{\R^n} x_i(x-m) e^{-\delta |x-m|^2/2}\rho(x) dx  }{\int_{\R^n} e^{-\delta |x-m|^2/2}\rho(x) dx }  -   \delta \frac{\left( \int_{\R^n} x_i  e^{-\delta |x-m|^2/2}\rho(x) dx \right)\left( \int_{\R^n} (x-m) e^{-\delta |x-m|^2/2}\rho(x) dx \right) }{\left(\int_{\R^n} e^{-\delta |x-m|^2/2}\rho(x) dx\right)^2 }\\
&=
 \delta \frac{\int_{\R^n} x_i x e^{-\delta |x-m|^2/2}\rho(x) dx  }{\int_{\R^n} e^{-\delta |x-m|^2/2}\rho(x) dx } -   \delta \frac{\left( \int_{\R^n} x_i  e^{-\delta |x-m|^2/2}\rho(x) dx \right)\left( \int_{\R^n} x e^{-\delta |x-m|^2/2}\rho(x) dx \right) }{\left(\int_{\R^n} e^{-\delta |x-m|^2/2}\rho(x) dx\right)^2 }.
\end{align*}
 
Hence,  the Jacobian of $T_{\delta}$ has entries $[D T_{\delta}(m)]_{ij} = \delta \operatorname{Cov}_{\mu_{m,\delta}}(x_i,x_j)$. Recalling the variance inequality above, 
$$
\| D T_{\delta}(m)\|_{op} = \delta \sup_{\sigma \in S^{d-1}} \Var_{\mu_{m,\delta}}(x\mapsto \sigma \cdot x) \leq \lambda_{\delta} < 1,
$$
so that $T_{\delta}$ is a contraction as claimed.   Hence, the desired existence and uniqueness of $m_{\delta}$ follows from the Banach Fixed Point Theorem. 
 
To prove the second claim, note that for any $m \neq m_{\delta}$ and $t\in[0,1)$, 
\begin{align*}
\frac{d}{dt} \int_{\R^n} e^{-\delta |x-((1-t)m + t m_{\delta}) |^2/2}\rho(x) dx &= \delta  \int_{\R^n} \!\!\langle x-((1-t)m + t m_{\delta}),m_{\delta}-m\rangle e^{-\delta |x-t m_{\delta} |^2/2}\rho(x) dx \\
&\propto \langle T_{\delta}\left( (1-t)m + t m_{\delta}  \right)  - ((1-t)m + t m_{\delta}),m_{\delta}-m\rangle  \\
&= \langle T_{\delta}\left( (1-t)m + t m_{\delta}  \right) -T_{\delta}(m_{\delta})  ,m_{\delta}-m\rangle + (1- t) |m_{\delta}-m |^2\\
&\geq -|T_{\delta}\left( (1-t)m + t m_{\delta}  \right) -T_{\delta}(m_{\delta})||m_{\delta}-m| + (1- t) |m_{\delta}-m |^2\\
& >  -|  (1-t)m - (1-t) m_{\delta}||m_{\delta}-m| + (1- t) |m_{\delta}-m |^2\\
&=0.
\end{align*}
The strict inequality holds since $T$ is a contraction and $(1-t)m + t m_{\delta}  \neq m_{\delta}$ for $t\in [0,1)$.  Thus, for any $m\in \R^n$ not equal to $m_{\delta}$, the map $t\mapsto \int_{\R^n} e^{-\delta |x-((1-t)m + t m_{\delta}) |^2/2}\rho(x) dx$ is strictly increasing on $[0,1)$, so that    $m \mapsto \int_{\R^n} e^{-\delta |x-m |^2/2}\rho(x) dx$ achieves a unique global maximum at $m_{\delta}$ as claimed.   
\par\vspace{3mm}
Toward proving the third claim, we first note that (ii) proved above yields a  uniform bound on $|m_{\delta}|$ for all $\delta>0$.  In particular, 
\begin{align*}
\int_{\R^n} |x| \rho(x) dx &\geq \int_{\R^n} |x| e^{-\delta |x-m_{\delta} |^2/2} \rho(x) dx\\
&\geq \left| \int_{\R^n} x e^{-\delta |x-m_{\delta} |^2/2} \rho(x) dx \right|\\
&=|m_{\delta}| \int_{\R^n} e^{-\delta |x-m_{\delta} |^2/2} \rho(x) dx\\
&\geq |m_{\delta}| \int_{\R^n} e^{-\delta |x |^2/2} \rho(x) dx\\
&\geq  |m_{\delta}|  \exp\left({- \tfrac{1}{2}\delta  \int_{\R^n} |x |^2 \rho(x) dx} \right).
\end{align*}
Since $\rho$ is log-concave, it has finite moments of all orders, and we conclude
$$
|m_{\delta}|\leq  \exp\left({ \tfrac{1}{2}\delta  \int_{\R^n} |x |^2 \rho(x) dx} \right)  \int_{\R^n} |x| \rho(x) dx <\infty.
$$
\par\vspace{3mm}
For each $\delta>0$, we introduce the more convenient notation $\mu_{\delta} = \mu_{m_{\delta}, \delta}$, where $m_{\delta}$ is defined as in part (i). By Taylor's theorem 
$$|e^{\epsilon f}-(1+\epsilon f)| \leq \frac{\epsilon^2 f^2}{2}e^{|\epsilon f|},$$
so it follows that 
\begin{align*}
\frac{\int_{\R^n} x e^{-V(x) - \delta|x-m_{\delta+{\epsilon}}|^2/2}  }{\int_{\R^n}  e^{-V(x) - (\delta+\epsilon) |x-m_{\delta+{\epsilon}} |^2/2} } &= \int_{\R^n} x \Big(1+\frac{\epsilon}{2}|x-m_{\delta+\epsilon}|^2 + O( \epsilon^2 |x-m_{\delta+\epsilon}|^4 e^{|\epsilon| |x-m_{\delta +\epsilon }|^2/2 } ) \Big) d\mu_{\delta+\epsilon} ,
\end{align*}
where the big-$O$ term hides only numerical constants. To show that the error term remains small after integration, note that 
\begin{align*}
&\left| \int_{\R^n} \left( x  \epsilon^2 |x-m_{\delta+\epsilon}|^4 e^{|\epsilon| |x-m_{\delta+\epsilon}|^2/2 }   \right) d\mu_{\delta+\epsilon}\right| \\
&\leq \epsilon^2\int_{\R^n} \left( |x|   |x-m_{\delta+\epsilon}|^4 e^{|\epsilon| |x-m_{\delta+\epsilon}|^2/2 }   \right) d\mu_{\delta+\epsilon} \\
&\leq  \epsilon^2\left( \int_{\R^n}   |x|^2   |x-m_{\delta+\epsilon}|^8 d\mu_{\delta+\epsilon}  \right)^{1/2}    \left( \int_{\R^n}  e^{|\epsilon| |x-m_{\delta+\epsilon}|^2 }   d\mu_{\delta+\epsilon}  \right)^{1/2}.
\end{align*}
Since $\mu_{\delta+\epsilon}$ is log-concave, an inequality of Borell ensures that 
$$
\left( \int_{\R^n} |x-m|^p d \mu_{\delta+\epsilon} \right)^{1/p} \leq C \frac{p}{q} \left( \int_{\R^n} |x-m|^q d \mu_{\delta+\epsilon} \right)^{1/q}
$$
for all $1\leq q \leq p <\infty$ and $m\in \R^n$, where $C$ is an absolute constant.   Thus, since $\int_{\R^n} |x - m_{\delta+\epsilon}  |^2 d \mu_{\delta+\epsilon} \leq \frac{n}{\delta+\epsilon}$ by the Brascamp-Lieb inequality and $\delta\mapsto |m_{\delta}|$ is bounded for $\delta>0$, the first term on the RHS involving polynomial moments is finite and  uniformly bounded (in terms of $\delta$) for all $\epsilon$ sufficiently small.  Additionally, since $\mu_{\delta+\epsilon}$ is  uniformly log-concave by construction, it satisfies $\mathrm{LSI}(\delta/2)$ for all $|\epsilon| < \delta/2$.  Hence, $\int_{\R^n}  e^{|\epsilon| |x-m_{\delta+\epsilon}|^2 }   d\mu_{\delta+\epsilon} $ is finite, and uniformly bounded in terms of $d,\delta$, for all $\epsilon$ sufficiently small.

Summarizing, we have 
\begin{align*}
\frac{\int_{\R^n} x e^{-V(x) - \delta|x-m_{\delta+{\epsilon}}|^2/2}  }{\int_{\R^n}  e^{-V(x) - (\delta+\epsilon) |x-m_{\delta+{\epsilon}} |^2/2} }  &=  \int_{\R^n} x \left(1+\frac{\epsilon}{2}|x-m_{\delta+\epsilon}|^2 \right) d\mu_{\delta+\epsilon} +O(\epsilon^2)\\
 &= m_{\delta+\epsilon} + \frac{\epsilon}{2}\int x|x-m_{\delta+\epsilon}|^2  d\mu_{\delta+\epsilon} +O(\epsilon^2)
\end{align*}
and, by similar arguments, 
\begin{align*}
\frac{\int_{\R^n}  e^{-V(x) - \delta|x-m_{\delta+{\epsilon}}|^2/2}  }{\int_{\R^n}  e^{-V(x) - (\delta+\epsilon) |x-m_{\delta+{\epsilon}} |^2/2} }&=   \int_{\R^n} \left(1+\frac{\epsilon}{2}|x-m_{\delta+\epsilon}|^2  \right) d\mu_{\delta+\epsilon} +O(\epsilon^2) \\
&= 1 + \frac{\epsilon}{2}\int |x-m_{\delta+\epsilon}|^2  d\mu_{\delta+\epsilon} +O(\epsilon^2),
\end{align*}
where the big-O terms hide finite constants that depend on ${\delta}$, but not on $\epsilon$. In particular, since
$$
T_{\delta}(m_{\delta + \epsilon})-m_{\delta + \epsilon} 
= \frac{\int_{\R^n} x e^{-V(x) - \delta|x-m_{\delta + \epsilon}|^2/2}  }{\int_{\R^n}  e^{-V(x) - \delta |x-m_{\delta + \epsilon}|^2/2} } -m_{\delta + \epsilon},
$$
we can conclude using the above estimates and uniform boundedness of $\delta\mapsto |m_{\delta}|$ that, for $\epsilon$ sufficiently small, 
$$
|T_{\delta}(m_{\delta + \epsilon})-m_{\delta + \epsilon}   | \leq  |\epsilon| C_{\delta},
$$
where $C_{\delta}<\infty$ depends  on ${\delta}$, but not $\epsilon$.

Now,  applying the second part of the Banach fixed-point theorem, we find for all $\epsilon$ sufficiently small  
\begin{align*}
|m_{\delta+\epsilon}-m_{\delta}| 
&\leq \frac{1}{1-\|T_{\delta}\|_{\operatorname{Lip}}} \left| T_{\delta}(m_{\delta + \epsilon})-m_{\delta + \epsilon}  \right| \leq \frac{|\epsilon| C_{\delta}}{1-\|T_{\delta}\|_{\operatorname{Lip}}},
\end{align*}
where we used the fact that $m_{\delta}$ is the fixed point of $T_{\delta}$.   Since $\|T_{\delta}\|_{\operatorname{Lip}}<1$ from the proof of (i), the proof of (iii) is complete.

Now, we proceed to establish claim (iv). For convenience, define for $\delta\geq 0$
$$
g(\delta):=-\log\left( \int_{\R^n} e^{- \delta |x-m_{\delta}|^2/2  } \rho(x)dx\right). 
$$
Since $\rho$ is a probability density, we have $g(0)=0$, so we focus henceforth on $\delta>0$.  
For $\delta', \delta>0$ the bound $ |x-m_{\delta'}|^2 \leq  |x-m_{\delta}|^2 + 2 \langle x-m_{\delta'},m_{\delta}-m_{\delta'} \rangle$ applies to give
\begin{align*}
g(\delta')-g(\delta) &= -\log\left(\frac{  \int_{\R^n} e^{  - \delta' |x-m_{\delta'}|^2/2  }\rho(x)dx }{  \int_{\R^n} e^{  - \delta |x-m_{\delta}|^2/2  }\rho(x)dx }\right)\\
&\leq -\log\left( {  \int_{\R^n} e^{ - (\delta'-\delta) |x-m_{\delta}|^2/2-\delta' \langle x-m_{\delta'},m_{\delta}-m_{\delta'} \rangle  }d\mu_{\delta} }\right)\\
&\leq    \int_{\R^n} \Big( (\delta'-\delta) |x-m_{\delta}|^2/2+ \delta' \langle x-m_{\delta'},m_{\delta}-m_{\delta'} \rangle  \Big) d\mu_{\delta} \\
& =(\delta'-\delta) \left( \frac{1}{2}\int_{\R^n}|x-m_{\delta}|^2d\mu_{\delta} + \frac{\delta'}{\delta'-\delta} |m_{\delta}-m_{\delta'} |^2\right) , 
\end{align*}
where we used convexity of $t\mapsto -\log(t)$ in the second inequality, and the final equality  used  $\int_{\R^n} x d\mu_{\delta} = m_{\delta}$.   Switching the roles of $\delta,\delta'$, we have the reverse inequality
\begin{align*}
g(\delta')-g(\delta) \geq (\delta'-\delta) \left( \frac{1}{2}\int_{\R^n}|x-m_{\delta'}|^2d\mu_{\delta'} - \frac{\delta}{\delta'-\delta} |m_{\delta}-m_{\delta'} |^2\right).
\end{align*}
By (iii), it holds that $|m_{\delta}-m_{\delta'} |^2 \leq L_{\delta}^2 |\delta-\delta'|^2$ for $\delta'$ sufficiently close to $\delta$. Additionally, $\int_{\R^n}|x-m_{\delta}|^2d\mu_{\delta} \leq \frac{n}{\delta}$ for each $\delta>0$ by the Brascamp-Lieb inequality.  Thus,  for $\delta',\delta>0$, with $|\delta-\delta'|$ sufficiently small, 
$$
| g(\delta')-g(\delta) | \leq |\delta'-\delta|\left(  \frac{n}{2\delta} + \delta' L_{\delta}^2 |\delta-\delta'| \right).
$$
In particular, $g$ is continuous on $(0,\infty)$ with upper Dini derivative bounded by 
$$
g_+'(\delta):=\limsup_{\epsilon \to 0^+} \frac{g(\delta+\epsilon)-g(\delta) }{\epsilon} \leq \frac{n}{2\delta}.
$$
Hence, we have for $\delta\geq \delta_0>0$
\begin{align}
g(\delta) \leq  g(\delta_0) +   \int^{\delta}_{\delta_0} g_+'(s) ds &\leq  g(\delta_0) + \frac{n}{2} \int^{\delta}_{\delta_0} \frac{1}{s} ds  =   g(\delta_0)   + \frac{n}{2}\log\frac{\delta}{\delta_0}.\label{g0Bound}
\end{align}
By (ii) and convexity of $t\mapsto -\log t$, we have
\begin{align*}
g(\delta)   &= -\log\left( \int_{\R^n} e^{ - \delta |x-m_{\delta}|^2/2  } \rho(x) dx\right) \\
&\leq -\log\left( \int_{\R^n} e^{  - \delta |x-m_0|^2/2  } \rho(x)dx\right) \\
&\leq \frac{\delta}{2}\int_{\R^n} |x-m_0|^2 \rho(x)dx.
\end{align*}
Thus, $g(\delta_0)\leq \frac{\delta_0}{2} \Var(\rho)$ for all $\delta_0> 0$.  Substituting into \eqref{g0Bound} and optimizing over $\delta_0>0$, we conclude the desired upper bound.

\end{document}